\documentclass[]{interact}

\usepackage{epstopdf}
\usepackage[caption=false]{subfig}

\usepackage[T2A,T1]{fontenc}

\usepackage[longnamesfirst,sort]{natbib}
\bibpunct[, ]{(}{)}{;}{a}{,}{,}
\renewcommand\bibfont{\fontsize{10}{12}\selectfont}

\theoremstyle{plain}

\theoremstyle{definition}

\theoremstyle{remark}

\newcommand{\rd}{R\&D}

\begin{document}


\title{On the long-term impact of non-formal learning in particle physics}

\author{
\name{J\'ulia Kekel\'akov\'a\textsuperscript{a} and Boris Tom\'a\v{s}ik\textsuperscript{a,b}\thanks{Email: boris.tomasik@cvut.cz}}
\affil{\textsuperscript{a}Univerzita Mateja Bela, Fakulta pr\'irodn\'ych vied, Tajovsk\'eho 40, 97401 Bansk\'a Bystrica, Slovakia; \\
\textsuperscript{b}\v{C}esk\'e vysok\'e u\v{c}en\'i technick\'e v Praze, Fakulta jadern\'a a fyzik\'aln\v{e} in\v{z}en\'yrsk\'a, B\v{r}ehov\'a 7, 
11519 Praha 1, Czech Republic}
}

\maketitle

\begin{abstract}
Since 2005, the global flagship of outreach activities in high-energy 
physics have been 
the International Particle Physics Masterclasses. We report on a 
survey performed among the participants from Slovakia and the 
Czech Republic, where we have studied the impact of Masterclasses
on their further careers and their attitude towards science and 
especially particle physics. More than a half of our respondents 
does not work in science or research and development. However, 
most of them report positive shift in their attitude towards science. 
A positive nudge to physics career is indicated among  
those being open to such a possibility. 
\end{abstract}


\section{Introduction}

In various areas of active (physics) research, often a large gap exists between the cutting edge results and the knowledge taught at secondary 
schools or distributed among the general public. This gap is partially filled by non-formal learning and outreach activities. 

High-energy physics, a.k.a.\ particle physics,  has been traditionally communicated rather intensely also outside of the field, 
as there has often been demand for the knowledge by the public.
The reason is perhaps that HEP pursuits the most fundamental principles of our Universe, which is perceived as very exciting. 
Nevertheless, the appearance  of this research 
field and its relation to the public are also very important for its 
own sustainability. Firstly, the complexity of the problems and
the scale of HEP experiments require rather large numbers of top-qualified workforce which in longer-term 
perspective are recruited among the 
young people graduating from secondary schools. 
Secondly, particle physics belongs to those branches of science that eventually require building large and costly infrastructures. Hence, the support from the governments---representing the tax-paying citizens---is 
vital. 
Outreach is the part of the strategy that addresses these issues. Thus the motivation for outreach is on all sides.

Since 2005---the 100th anniversary of Einstein's Annus Mirabilis---one of the internationally most important outreach activities in HEP
has been developed: the International Particle Physics Masterclasses (IMC). Institutes from Slovakia and the Czech Republic have participated 
in this initiative since its beginning. 

Due to its general attraction and longevity one could expect that IMC have already influenced a generation of young people and 
motivated some of the careers towards HEP, or generally to STEM. Furthermore, it could have contributed to the perception of science
among those who chose non-STEM careers. 
These are interesting hypotheses and until now we have had no data against which they could have been confronted. 
In this study, we report on a survey that addresses these hypotheses and demonstrates the long-term impact of IMC. 

Our survey shows that while more participants chose careers outside science or research\&development (\rd), the majority maintains positive
attitude towards science. The recollections about IMC are mostly very positive.

We describe the International Particle Physics Masterclasses in the next Section. In Section \ref{s:data} we explain how data were collected
on which our survey is based. Results are presented in Section \ref{s:results} and the conclusions are summarised in Section \ref{s:conc}.


\section{International Particle Physics Masterclasses}
\label{s:IMC}

The principal idea that governs the Masterclasses is to provide the upper secondary level pupils the genuine experience 
of particle physics. The advertising mottos `Hands on CERN' and `Become a particle physicist for one day' give the first idea about its 
motivation and agenda. 
The event (almost) always takes place at a university or research institute that runs active research programme in particle physics. 
One of the aims is to also bring the pupils into the environment where research is performed. 

The agenda of the event is planned for one day and the highlight is an analysis of data collected by one of the major particle 
physics experiments. Participants thus get hands-on experience mimicking the work of particle physicists. 

No previous knowledge is necessary in order to participate in IMC. Therefore, the morning programme usually includes lectures
that provide the elementary knowledge about theoretical concepts of particle physics and explain the basic experimental techniques
used. 

After the lunch, the actual hands-on activity is introduced and explained.
Participants work on it individually or in pairs, and it takes about two hours. 
Currently, all four large experiments at the LHC (CERN) offer activities that are being used in Masterclasses. 
Some of the collaborations even prepared two different activities. 
In addition to CERN experiments there are 
exercises prepared by BELLE~II (KEK Laboratory, Japan), MINER$\nu$A (Fermi National Laboratory, USA), 
Pierre Auger (cosmic ray detection array, Argentina) and one activity prepared by GSI Darmstadt on particle therapy 
in oncology.

The international character of the research in particle physics is also illustrated in the subsequent videoconference. There, 
up to five institutes connect and discuss---together with moderators from a major particle physics laboratory---their results. The videoconference
is arranged by the global management of the event in such a way, that at all participating institutes the same hands-on activity 
has been performed. The participants in the videoconference may also pose questions about particle physics and related issues 
in general, which are answered by the moderators---who are professional physicists themselves. 

The hands-on activity is rather involved. Such an endeavor, together with the international coordination of the videoconferences
is only possible thanks to the joint effort managed by the International Particle Physics Outreach Group (IPPOG)\footnote{{https://ippog.org}}.
Formally, IPPOG is an international collaboration, hosted by CERN, and facilitating exchange of ideas as well as coordinating outreach 
activities in the participating countries and with the major particle physics laboratories and experiments. 

For the first time on international level, IMC were organised by European Particle Physics Outreach Group (EPPOG) in 2005. 
It was in 2010 that EPPOG evolved internationally and became IPPOG.
Today, each year about 13,000 pupils participate in IMC that take place at some 220 institutes from 55 countries. 

More details about IMC can be found in the literature e.g.\ by \cite{Kobel05,Foka13,Bilow14,Bilow22}, 
and the information about IPPOG can be obtained from their web-site.


\section{Data collection}
\label{s:data}

First survey which focussed on the performance of IMC was undertaken 
already in 2005 and reported by \cite{Kobel05}.

Both Czech Republic and Slovakia have been participating in IMC since its first edition in 2005. Between 2011 and 2015
surveys were performed at most of the participating institutions in Slovakia and once also at the Czech Technical University in Prague,
Czech Republic. Those surveys were mainly focused on the 
educational background of the participants and the assessment of the IMC 
with the aim to further develop and improve the event. Results were partly summarised by \cite{Beniacikova11,Tomasik13,Cecire17}
and/or used internally. 

The surveys were performed by means of answer sheets that were (usually) distributed and collected just after the end of 
IMC while the participants were still at the venue. In one case 
two questionnaires were collected---one before and one after the IMC---in order 
to directly measure the impact of IMC. The total number of collected sets of answers over all years is somewhat below 1000. The 
surveys were anonymous. Nevertheless, in the end we suggested that if the participants agreed, they could leave us with their email addresses 
for the purpose of a later survey. 

In April and May 2023 we have performed two new surveys, from which we present results in this paper. 

\paragraph*{Survey 1} has been performed among former participants of IMC, who  left their email addresses 
with us. In this way we have retrieved 484 email addresses from the archived answer sheets. 

We administered an anonymous 
online questionnaire that has been implemented with the help of google forms. In included 10 questions. 
We deliberately kept the form short in order not to discourage the participants from filling it in. The questions were in Slovak for participants 
from both Slovakia and Czech Republic. The questions, translated to English, are listed in Table~\ref{t:q1}.
%
%
\begin{table}[t]
\caption{Questions asked in the questionnaire of Survey 1.}
\centerline{
\begin{tabular}{| c | p{0.595\textwidth} | p{0.24\textwidth} |} 
\hline 
& Question & Mode of answer \\ 
\hline\hline
Q1.1 & In which year have you participated in Masterclasses? & multiple choice \\
\hline
Q1.2 & At which place have you participated in Masterclasses? & multiple choice \\ 
\hline
Q1.3 & I came to Masterclasses with the ambition to study particle or nuclear physics. 
& 5-point Likert scale \\
\hline
Q1.4 & Masterclasses have influenced my decision to study physics or related subject. & 5-point Likert scale \\
\hline
Q1.5 & I contribute to the development of science---I am scientific associate, or I work for a company focussed on R\&D, or 
I want to work in this field after I finish my study. & 5-point Likert scale \\
\hline
Q1.6 & The content of my current employment or study is: &  short free text  \\
\hline
Q1.7 & Thanks to Masterclasses I positively changed my opinion about science, research, and physics. &  free text  \\
\hline
Q1.8 & I am interested in the news from CERN and/or news from science and particle physics. & 5-point Likert scale \\
\hline
Q1.9 & If you want to leave us a message, you can type it here. & long free text \\
\hline
Q1.10 & This is the end of the questionnaire. If you agree with the comparison of your answers with the answers that you gave 
after the event, leave us your email address here. Thank you! & short free text \\
\hline
\end{tabular}
}
\label{t:q1}
\end{table}
%
%
The first two questions serve statistic purposes, mainly. Q1.3 aims at reconstruction of the motivation prior to experiencing IMC. 
Questions Q1.4--Q1.8 explore the actual impact of IMC as perceived today. We will mention the comments that we received in 
Q1.9 below, and look at the correlation between the previous survey and this one thanks to emails collected in Q1.10.

Out of the 484 invitations to our survey, 133 messages bounced back as undeliverable, hence we assume that 351 messages 
were delivered. From those, 71 participants filled our online questionnaire. This is the sample we work with. Furthermore, 36 people 
revealed their email addresses and we were able to connect their responses with the ones that were collected in the past. 

\paragraph*{Survey 2} was performed in parallel to Survey 1 with the aim to identify if and how  IMC influenced young researchers 
in HEP. In this case, we asked the senior faculty members and team leaders at relevant Czech and Slovak institutions to 
forward our invitation to the study to younger colleagues. The participation could not be enforced, thus the sample may be biased 
and possibly not all eligible people may have responded. We collected 19 responses in total. This number is to be compared 
with the relevant size of the community, which we estimate as 100 young HEP practitioners in the Czech Republic and 25 in Slovakia, 
totalling to 125. 
The numbers include young colleagues that---due to their age---could have participated in IMC and can be clearly assigned to 
research in HEP. This results in a group from master study level up to about 35 years of age. The numbers were estimated  after 
requesting the (approximate) headcount from team leaders at all relevant institutions. 

Since our surveys were anonymous, unless the respondents volunteered to 
reveal their identity, we do not have information about possible overlap between participants to Survey 1 and Survey 2. From anecdotal discussions we know that there is some overlap, but our method does not allow for a more detailed information. 

%
%
\begin{table}[t]
\caption{Questions asked in the questionnaire of Survey 2.}
\begin{tabular}{| c | p{0.595\textwidth} | p{0.24\textwidth} |} 
\hline 
& Question & Mode of answer \\ 
\hline\hline
Q2.1 & In which year have you participated in Masterclasses? & multiple choice \\
\hline
Q2.2 & At which place have you participated in Masterclasses? & multiple choice \\ 
\hline
Q2.3 & I came to Masterclasses with the ambition to study particle or nuclear physics. 
& 5-point Likert scale \\
\hline
Q2.4 & Masterclasses have influenced my decision to study particle physics. & 5-point Likert scale \\
\hline
Q2.5 & Later I have helped with the organisation of the Masterclasses at the institution where I studied or worked. 
& 5-point Likert scale \\
\hline
Q2.6 & I would like to help with the organisation of Masterclasses in the future. & 5-point Likert scale \\
\hline
Q2.7 & In particle or nuclear physics I am rather focussed on... & 
multiple choice from:\newline experiment; theory; \newline phenomenology; nothing yet, because I study
\\
\hline
Q2.8 & My current professional status could be described as... & 
radio buttons with the possibilities:\newline student (bachelor or master level);\newline doctoral student;\newline postdoc;\newline 
scientific staff;\newline university teacher; \newline other\\
\hline
Q2.9 & If you want to leave us with a message concerning Masterclasses, type it here. We will be grateful, e.g., for proposals 
what can be improved. In any case, thank you for filling the questionnaire.
& long free text \\
\hline
\end{tabular}

\label{t:q2}
\end{table}
%
%
The questions of Survey 2 are summarised in Table~\ref{t:q2}. Part of them is similar to Survey 1, but in case of Survey 2 we know 
that the answers are given by practitioners in HEP. The other part of the questions thus rather  aims  on the research focus of the participants
and the attitude to currently organised IMC. 

For brevity, below we will refer to the respondents of Survey 1 as \emph{fans}, while respondents of Survey 2 will be called \emph{practitioners}.


\section{Results}
\label{s:results}

%
%
\begin{figure}[h]
\centering
\includegraphics[width=0.9\textwidth]{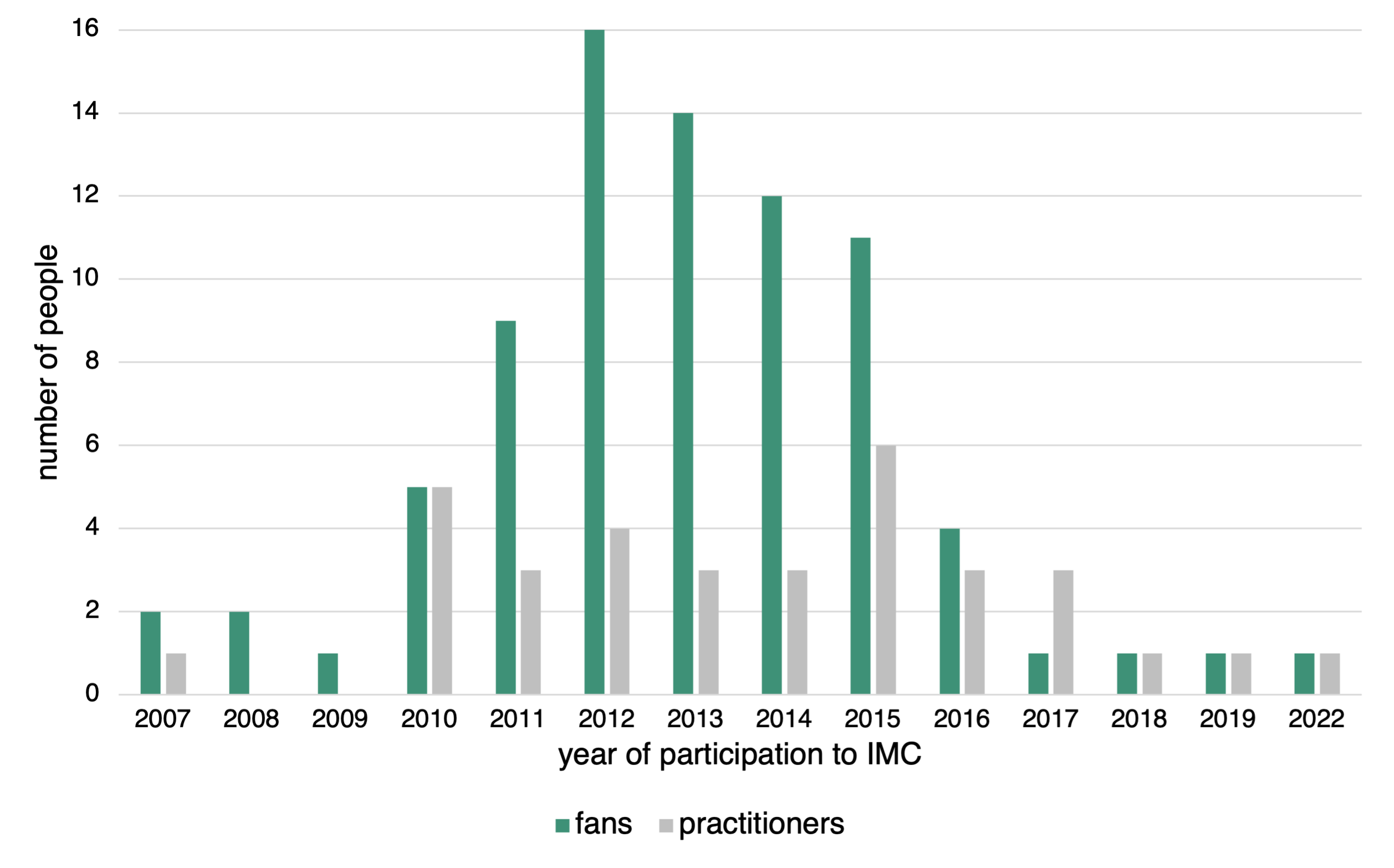}
\caption{Distribution of the years of participation to IMC for fans (dark columns) and practitioners (light columns).} \label{f:years}
\end{figure}
%
%
Histograms in Fig.~\ref{f:years} show the distribution of years when our respondents participated in IMC. Based on our data 
collection procedure it is not surprising that the fans mostly participated between 2011 and 2015, when the previous survey 
was performed. Individual cases outside of this time interval either indicate multiple participation to IMC by one respondent, 
or erroneous assignment of the year. There were 12 responses which could not recall the actual year and they are not included 
here. The distribution of years is wider for the practitioners. Participation after 2019 indicates the survey leaked also to 
current bachelor-level students. We decided to keep their responses in the sample, thus making it more informative. 

%
%
\begin{figure}[b!]
\centering
\includegraphics[width=0.9\textwidth]{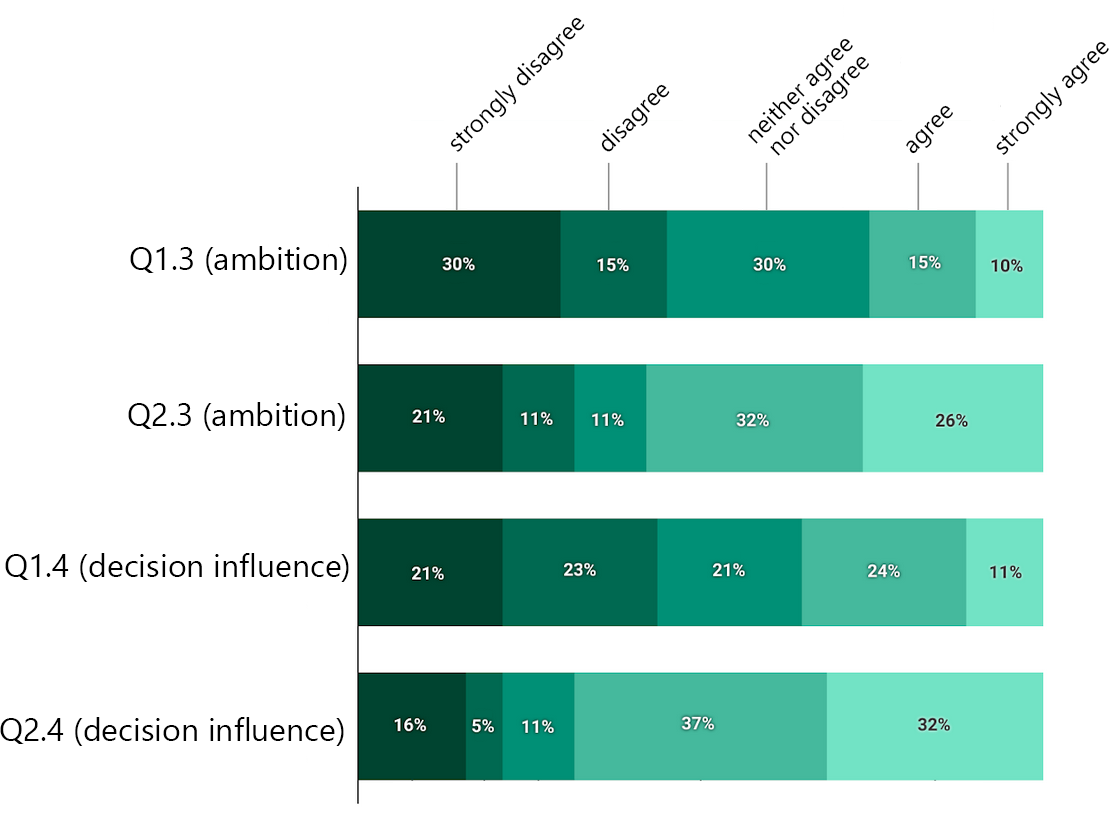}
\caption{Distributions of answers to questions (from top to bottom) Q1.3, Q2.3, Q1.4, Q2.4. } \label{f:bars-motiv}
\end{figure}
%
%
Next, we study to what extent IMC have directed our respondents towards a career in particle physics. In Figure~\ref{f:bars-motiv}
we combine the results from both surveys. In the general group of fans, people with the ambition to go to particle physics 
make up slight minority. Not surprisingly, more than a half of the practitioners were motivated to proceed to particle physics 
after they experienced IMC. 
Interestingly enough, according to Q2.3, 
6 out of 19 practitioners had no such ambition before coming to IMC. 
Since they did end up as particle physicists, this means that they have changed their minds later. The impact of IMC
in directing towards particle physics is also quite different in the two groups. While practitioners were clearly influenced, 
slight majority of the fans feel rather not influenced.

%
%
\begin{figure}[t]
\centering
\includegraphics[width=0.5\textwidth]{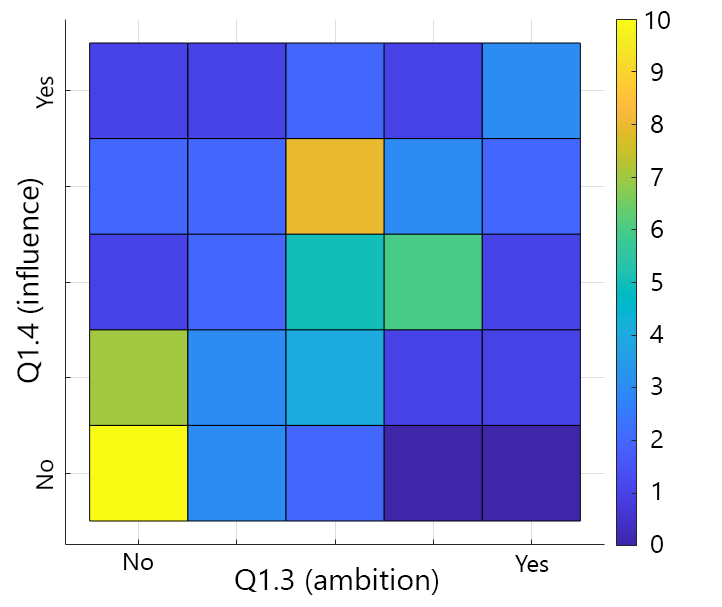}
\caption{Two-dimemnsional histogram of answers to questions Q1.3 (ambition) and Q1.4 (influence) for the sample of fans} 
\label{f:amb-mot}
\end{figure}
%
%
In order to understand the motivating effect better, we look in Figure~\ref{f:amb-mot} into the correlation of answers to 
Q1.3 and Q1.4, i.e., questions about the pre-existing ambition to study particle physics and the influence of IMC on personal 
motivation, for the sample of fans. The histogram shows a clear peak for people neither planning nor being influenced towards 
particle physics. Nevertheless, on the 'yes'-side of the histogram we see a hint of anti-correlation between the ambition 
(Q1.3) and decision influence (Q1.4): people with ambivalent attitude towards particle physics seem to be nudged towards it
while those with positive attitude report no additional influence. 

There are not enough data to produce a similar plot for the practitioners, hence we refrain from it. 

%
%
\begin{figure}[t]
\centering
\includegraphics[width=1\textwidth]{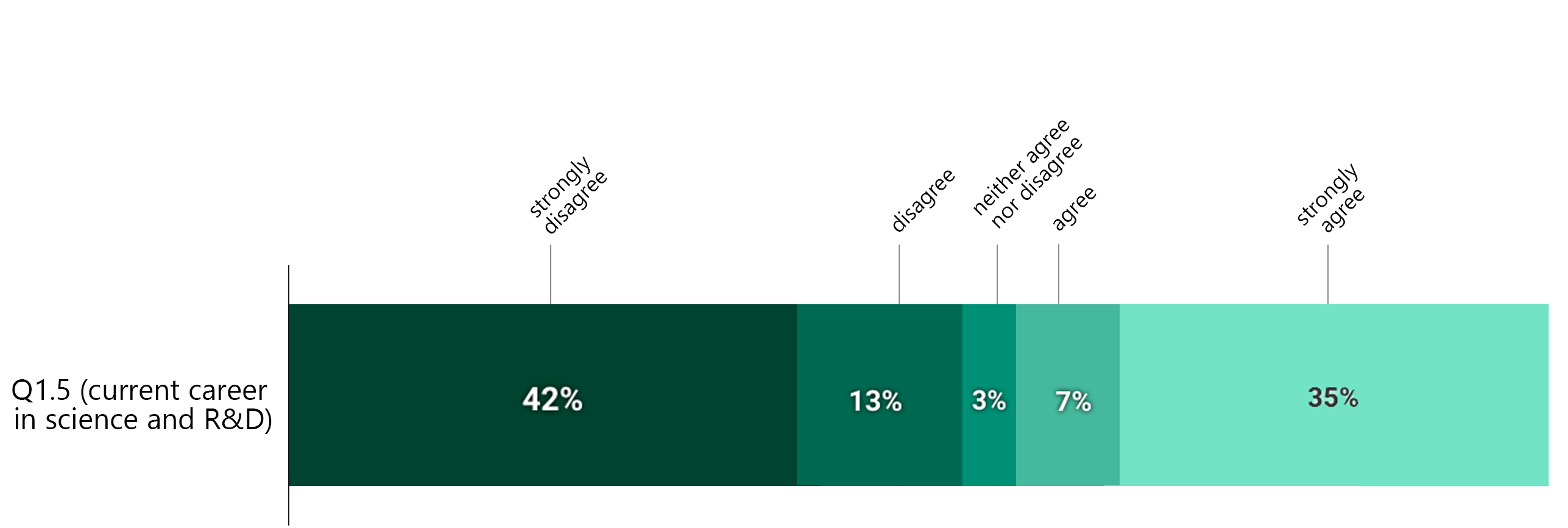}
\caption{Distribution of the answers to Q1.5: current career in science and R\&D.} 
\label{f:careers}
\end{figure}
\begin{figure}
\centering
\includegraphics[width=0.95\textwidth]{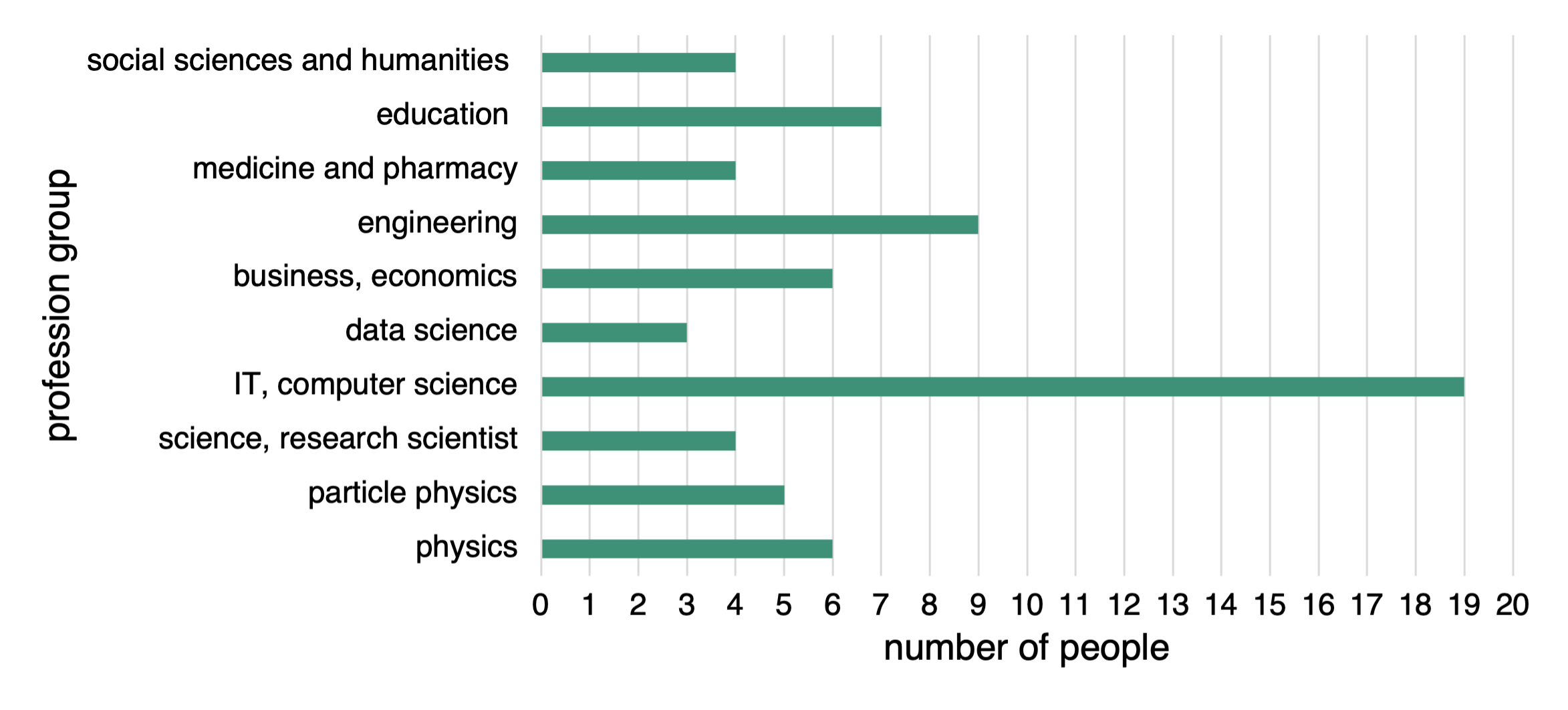}
\caption{Current distribution of professions within the sample of fans.} 
\label{f:professio}
\end{figure}

%
%
We were interested in further evolution of the careers of former IMC participants. As can be seen in Figure~\ref{f:careers},
slightly more than a half of the fans actually work outside science and R\&D. A more detailed information is summarised in 
Figure~\ref{f:professio}.
%
%
%
%
By far the most populated profession group is related to IT. Another prominent group can be identified if we put 
together all healthcare-related professions. The education group mostly includes science teachers. In general, STEM-related 
careers prevail, with only 10 out of 71 respondents falling clearly out of this field, i.e., to social sciences, humanities, economy, 
or business. 

\begin{figure}
\centering
\includegraphics[width=0.7\textwidth]{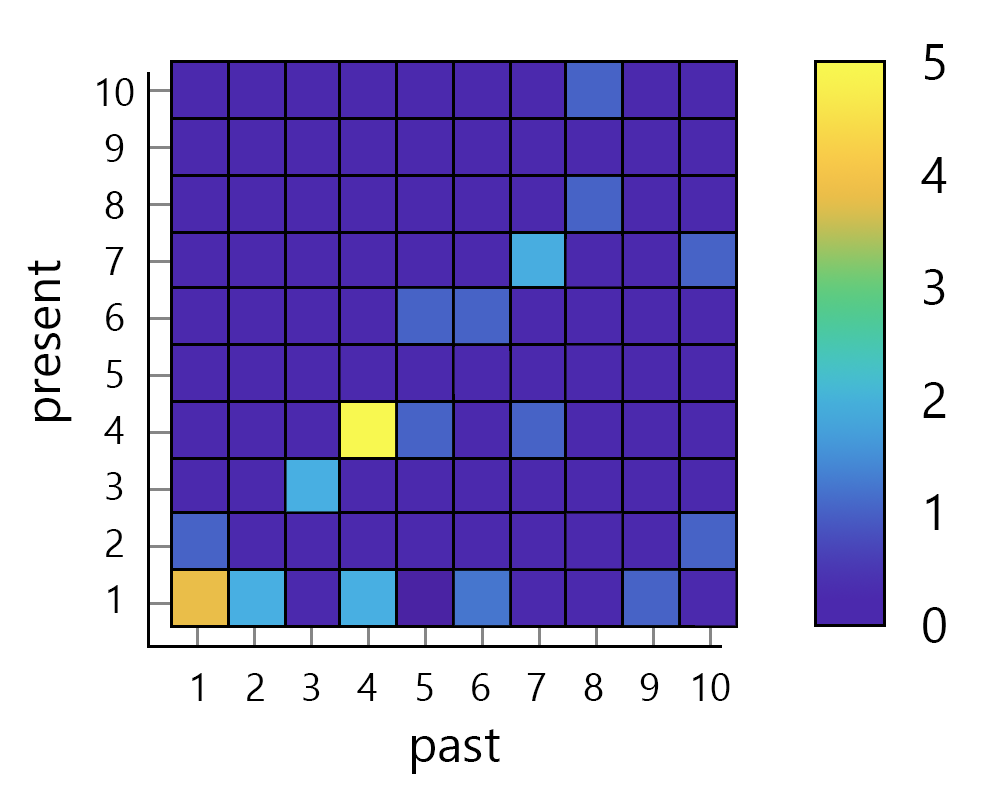}
\caption{Histogram of career plans immediately after IMC vs.\ current reality. Coding of the profession groups: 1 physics; 2 particle physics;
3 science, research scientist; 4 IT, computer science; 5 data science; 6 business, economics; 7 engineering; 8 medicine and pharmacy;
9 education; 10 social sciences and humanities.} 
\label{f:car-corr}
\end{figure}

Linking the current responses with those collected about ten years ago allows us to gain some insight into the evolution of 
career plans. The correlation of previous plans with current reality is analysed in Figure~\ref{f:car-corr}.
%
%
%
%
It is based on the 36 responses that could have been linked. We divided all professions into 10 groups, see the caption. 
If there was no change of plans, data would be aligned along the diagonal. Such a trend is roughly visible, with two pronounced
peaks; the main for IT and the next for  physics. In addition to the diagonal there are hints of two more effects. Firstly, 
there is a group of respondents who originally planned various careers but ended up with  physics. Secondly, 
a similar, though smaller group ended up finally in IT.

%
%
\begin{figure}
\centering
\includegraphics[width=0.9\textwidth]{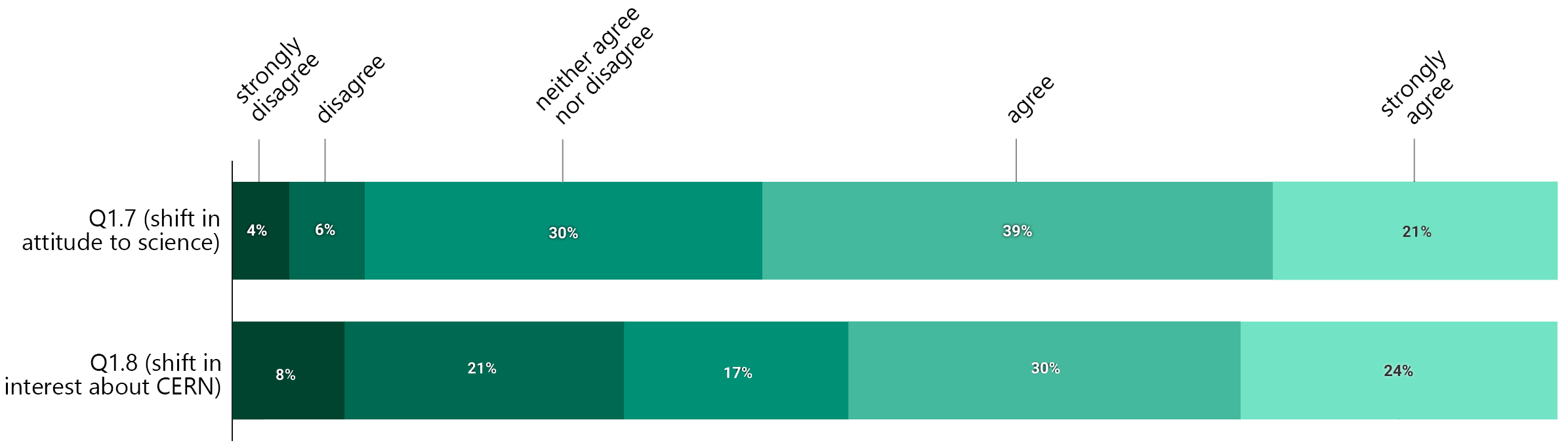}
\caption{Distributions of answers to questions Q1.7 (shift in attitude to science) and Q1.8 (shift in interest about CERN).} 
\label{f:attid}
\end{figure}
%
%
Another focus of our surveys was  in the perception of particle physics and science, in general. 
We summarise in Figure~\ref{f:attid} the answers to questions Q1.7, and Q1.8, which aim at the positive change 
in the attitude to science and research, and the interest in CERN and particle physics in general, respectively. 
In case of the attitude we see a positive effect of the IMC. In case of the interest in particle physics this is less pronounced. 

%
%
\begin{figure}
\centering

\includegraphics[width=1.0\textwidth]{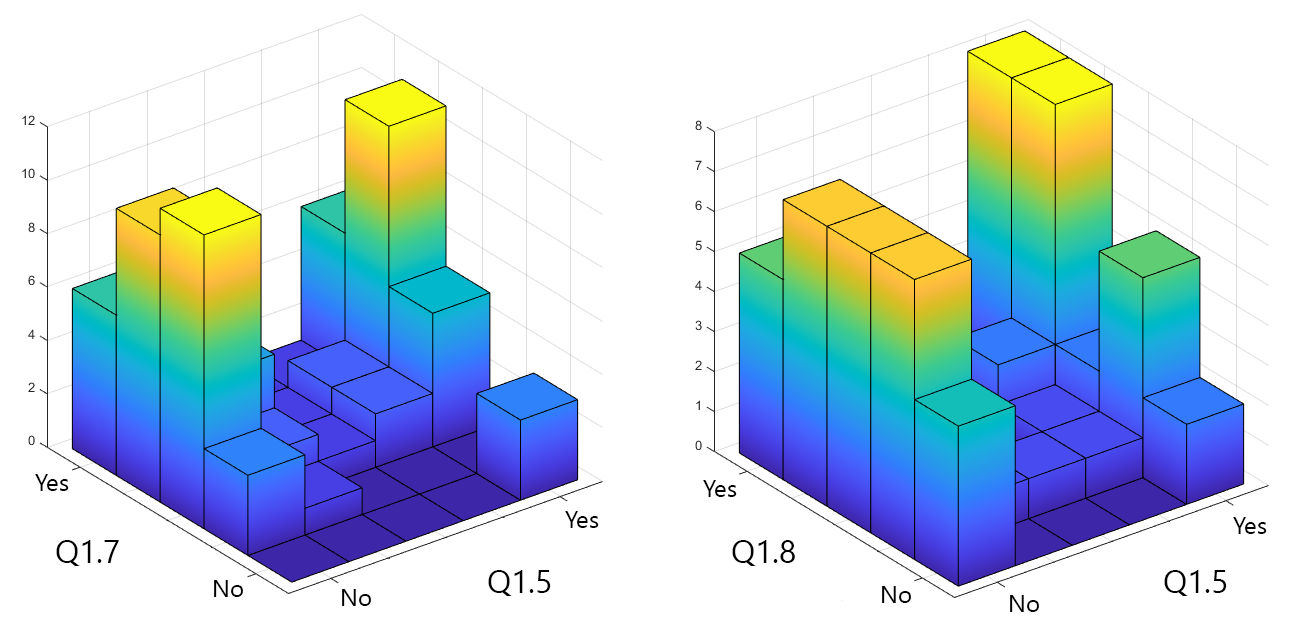}
\caption{Two dimensional histograms of the combinations of answers to Q1.7 with Q1.5 (attitude to science and career in science or R\&D, left)
and Q1.8 with Q1.5 (interest in news from CERN and career in science or R\&D, right). }
\label{f:att-corr}
\end{figure}
%
%
It is interesting to see how the answers to Q1.7 and Q1.8 depend on the professional orientation of the respondent. To this end, 
we show in Figure~\ref{f:att-corr} two-dimensional histogram that combines answers to these questions with those to Q1.5. 
The histograms demonstrate that both the positive shift in attitude to science and the interest in CERN and particle physics 
are more pronounced with respondents working in science or R\&D. 

Survey 2 also clearly showed that the practitioners who participated in IMC are very happy to help with IMC at their institutes. 
Most of them work on experiments, in comparison  to theory or phenomenology, which is usual distribution in HEP. 

We would like to close this section with the messages that the respondents left us as free text. They unanimously evaluated IMC very positively
even if they decided to choose a different career path. Some of these decisions were motivated by the financial attractiveness of the IT sector. 

These answers illustrate that there seems to be a more profound idea that is communicated indirectly in the IMC events. 
It is related to the meaningfulness of the devoted (scientific) work, which is surely transferable beyond particle physics and is much 
more general asset in life. 

Since it may be interesting, where possible we identified the gender of the respondent, which in Slovak language can be inferred from 
the form of the verbs used in the answers. There were 5 responses from males and 7 from females, while 8 responses could not be uniquely 
assigned. We did not collect the data on gender because we were not interested in this aspect. Nevertheless, the observed distribution 
is in agreement with our experience that the participants to IMC are rather balanced. 

Selected messages follow: 

\begin{itemize}
\item 
'Very good event. (IT pays better.)`\\
('Ve\v{l}mi dobr\'e podujatie. (IT sektor platí lep\v{s}ie.)`)
\item
'As a teacher I would recommend it to pupils interested in physics.`\\ 
('Ako učiteľka by som to odporučila absolvovať žiakom so záujmom o fyziku.`)
\item 
'I participated in Masterclasses 12 years ago and I still recollect it very well. It was tremendous experience for me as a student who was interested in physics and planned to study this specialisation. It showed me new possibilities and I gained new encouragement to proceed this way also to the university. I am thankful for this opportunity.`\\
('Podujatie Masterclasses som absolvovala pred 12timi rokmi a doteraz si na to veľmi dobre spomínam. Bola to úžasná skúsenosť pre mňa ako študentku, ktorá sa zaujímala o fyziku a mala v pláne ist študovať tento odbor. Ukázalo mi to nové možnosti a nabrala som ešte väčšie odhodlanie ísť touto cestou aj na vysokej škole. Som vďačná za túto príležitosť.`)
\item 
'In my opinion, Masterclasses are a superb event. I would not say that it changed my opinion about science; I held it positive already before. It gave me kind of a first practical contact with science, how it works, what is the state of the art and what are open questions in the given area. It also helped me to grasp some things, that we learned in secondary school, because perhaps even our teachers did not understand them so deeply, so that they could present it so simply and logically. It was a contact with the fact that if you pursue something more deeply, it could make sense. I think that Masterclasses make sense, even if you finally decide to study something else than particle physics. Today, many projects are interdisciplinary and then it is an advantage if you had the possibility to get a flavor of the  terminology and get some basic knowledge also from different specialisations.`\\
('Masterclasses je pod\v{l}a m\v{n}a super podujatie. Nepovedala by som, \v{z}e by mi to zmenilo n\'azor na vedu, ten som mala u\v{z} aj pred t\'ym pozit\'ivny. Dalo mi to tak\'y prv\'y praktick\'y kontakt s vedou, ako prebieha a \v{c}o state of the art a \v{c}o s\'u open questions v danej oblasti. Tie\v{z} mi to pomohlo pochopi\v{t} niektor\'e veci, pochopi\v{t}, ktor\'e sme sa u\v{c}ili na strednej, lebo asi ani na\v{s}i u\v{c}itelia tomu tak do h\'lbky nerozumeli, aby to vedeli jednoducho a logicky poda\v{t}. Bol to kontakt s t\'ym, \v{z}e ke\v{d} sa \v{c}lovek nie\v{c}omu venuje trochu viac do h\'lbky, m\^o\v{z}e to d\'ava\v{t} zmysel. Mysl\'im si, \v{z}e projekt Masterclasses m\'a zmysel, aj ke\v{d} sa \v{c}lovek nakoniec rozhodne robi\v{t} a \v{s}tudova\v{t} nie\v{c}o in\'e ako \v{c}asticov\'u fyziku. Ve\v{l}a projektov je teraz medziodborov\'ych a tam je v\'yhodou, ak m\'a \v{c}lovek mo\v{z}nos\v{t} na\v{c}uchn\'u\v{t} aj do terminol\'ogie a z\'iska\v{t} nejak\'e z\'akladn\'e vedomosti aj z in\'ych odborov.`)
\item 
'I would like to praise the organisation of this event. During my study I participated in such events joyfully and often, and this one particularly stayed in my memory, because it seems to me that there I learned in a short time a lot, and it certainly increased my interest for this field, even though later---perhaps rather due to practical reasons---I decided to study informatics. Thank you!\hspace{0pt}`\\
('Chcem pochv\'ali\v{t} organiz\'aciu tohoto podujatia. Po\v{c}as \v{s}t\'udia som sa podobn\'ych akci\'i z\'u\v{c}ast\v{n}ovala rada a \v{c}asto, a tento mi obzvl\'a\v{s}\v{t} utkvel v pam\"ati, lebo m\'am pocit , \v{z}e som sa tam za kr\'atky \v{c}as nau\v{c}ila ve\v{l}a a ur\v{c}ite to v tom \v{c}ase zv\'y\v{s}ilo m\^oj z\'aujem o tento obor, hoci som sa nesk\^or, mo\v{z}no viac z praktick\'ych d\^ovodov, rozhodla pre \v{s}t\'udium informatiky. V\v{d}aka!\hspace{0pt}`)
\item
'Even though it was a long time ago, I have very good recollections of Masterclasses and I think that such activities are of great significance. Even if I did not study physics at last, but informatics, CERN still bugged me and finally I got there as a fellow and spent three years in the IT division. Just by chance, right now I am sitting in a train to Geneva on my way to visit my former colleagues. :)` \\
('Aj keď to už bolo dávno, na Masterclasses mám veľmi dobré spomienky a myslím si, že takéto aktivity majú veľký význam. Aj keď som nakoniec nešla študovať fyziku, ale informatiku, CERN mi zostal ako chrobák v hlave a nakoniec som sa tam dostala na fellowship a strávila som tri roky v IT oddelení. Zhodou okolností práve sedím vo vlaku do Ženevy a idem pozrieť bývalých kolegov :)`)
\end{itemize}

\section{Conclusions}
\label{s:conc}

We found an indication in our data that International Particle Physics Masterclasses do have a nudging effect on participants 
who do not exclude the possibility of becoming a scientist or even particle physicist. 
They appear as very effective learning environment where in short time great experience is provided to those who participate. 
 
 Nevertheless, a larger portion of former participants does not pursue scientific career, or a career in R\&D. IMC then provide a kind of 
 cultural transfer in which the (good) practices followed in particle physics are exported to other fields and professions. In line
 with this, IMC contributes to the generally positive acceptance of science. 
 
 The recruiting function is often stressed and even mentioned frequently in an IMC event. ('You can become particle physicist and 
 solve these open problems!\hspace{0pt}`)
 Nevertheless, it should be acknowledged and appreciated that perhaps more important impact of IMC is in the export of knowledge 
 and culture beyond particle physics and enriching the future generation of citizens.


\section*{Acknowledgements}

We thank Ivan Melo and Vojt\v{e}ch Pleskot for insightful discussions that helped us in performing the reported survey, as well as critical reading and comments to the manuscript. 
We are grateful to the colleagues who provided inputs and forwarded our requests to participate in Survey 2: Pavol Barto\v{s}, Jaroslav Biel\v{c}\'ik,
Marek Bombara, Peter Chochula, Michal Mere\v{s}, Vojt\v{e}ch Pleskot, Marek Ta\v{s}evsk\'y, Martin Venhart.
We also thank the students who helped us with processing data from the answer sheets: Nat\'alia Dzia{\l}ak and Dominika Kru\v{z}\'ikov\'a. 


\section*{Funding}

IMC and this study have been supported by the Ministry of Education, Science, Research and Sports of the Slovak Republic.


\end{document}